\documentclass{aa}
\usepackage[varg]{txfonts}
\usepackage{natbib, aas_macros}
\usepackage{graphicx} 
\usepackage{amsmath}
\usepackage{lineno}
\usepackage{floatrow}
\begin{document}

\title{Interior structures and tidal heating in the TRAPPIST-1 planets}

\author{Amy C. Barr\inst{1}
 \and Vera Dobos\inst{2,3,4}
 \and L\'aszl\'o L. Kiss\inst{2,5}}


\offprints{A. C. Barr, \email{amy@psi.edu}}

\institute{Planetary Science Institute, 1700 E. Ft. Lowell, Suite 106, Tucson, AZ 85719, USA
  \and Konkoly Thege Mikl\'os Astronomical Institute, Research Centre for Astronomy and Earth Sciences, Hungarian Academy of Sciences, H--1121 Konkoly Thege Mikl\'os \'ut 15-17, Budapest, Hungary
  \and Geodetic and Geophysical Institute, Research Centre for Astronomy and Earth Sciences, Hungarian Academy of Sciences, H--9400 Csatkai Endre u. 6-8., Sopron, Hungary
  \and
ELTE E\"otv\"os Lor\'and University, Gothard Astrophysical Observatory, Szombathely, Szent Imre h. u. 112, Hungary
  \and
Sydney Institute for Astronomy, School of Physics A28, University of Sydney, NSW 2006, Australia}

\date{Received 26 September 2017 / Accepted 14 December 2017}

\abstract
  {With seven planets, the TRAPPIST-1 system has among the largest number of exoplanets discovered in a single system so far.  The system is of astrobiological interest, because three of its planets orbit in the habitable zone of the ultracool M dwarf.}
  {We aim to determine interior structures for each planet and estimate the temperatures of their rock mantles due to a balance between tidal heating and convective heat transport to assess their habitability. We also aim to determine the precision in mass and radius necessary to determine the planets' compositions.}
    {Assuming the planets are composed of uniform-density noncompressible materials (iron, rock, H$_2$O), we determine possible compositional models and interior structures for each planet. We also construct a tidal heat generation model using a single uniform viscosity and rigidity based on each planet's composition.}
  {The compositions for planets b, c, d, and e remain uncertain given the error bars on mass and radius. With the exception of TRAPPIST-1c, all have densities low enough to indicate the presence of significant H$_2$O.  Planets b and c experience enough heating from planetary tides to maintain magma oceans in their rock mantles; planet c may have surface eruptions of silicate magma, potentially detectable with next-generation instrumentation. Tidal heat fluxes on planets d, e, and f are twenty times higher than Earth's mean heat flow.}
 {Planets d and e are the most likely to be habitable. Planet d avoids the runaway greenhouse state if its albedo is $\gtrsim$ 0.3.  Determining the planet's masses within $\sim$~0.1 -- 0.5 Earth masses would confirm or rule out the presence of H$_2$O and/or iron. Understanding the geodynamics of  ice-rich planets f, g, and h requires more sophisticated modeling that can self-consistently balance heat production and transport in both rock and ice layers.}
\maketitle

\section{Introduction}

The recent discovery of seven roughly Earth-sized planets orbiting the low-mass star TRAPPIST-1 \citep{gillon17} has vaulted this system to the forefront of exoplanetary characterization.  The planets orbit the star with semi-major axes $< 0.1$~AU, and orbital periods of a few Earth days.  Given their proximity to the star, and the star's low mass and luminosity, the surface of each planet has a moderate temperature, ranging from $\sim 160$ to 400 K \citep{gillon17}.  Given the planets' mean densities, these low temperatures suggest that some might have solid surfaces composed of H$_2$O ice and/or rock.  The planets also have non-zero orbital eccentricities \citep{gillon17, wang17}, so that the tidal forces on the bodies vary with time, resulting in heating of their interiors by tidal dissipation \citep{luger17}.  Tidal heating is an important energy source in the satellites of the outer planets in our solar system \citep[e.g.,][]{mcewen04, greeley04, spencer09} and could significantly enhance the habitable volume in the galaxy by warming exoplanets and their satellites (exomoons) \citep{dobos15}.

Here, ``tidal dissipation'' (or tidal heating) refers to the process of the dissipation of orbital energy in the interior of a solid body (a moon or a planet). For a single secondary object, that is, a lone moon orbiting a planet, or a lone planet orbiting a star, tidal dissipation rapidly decreases the orbital eccentricity of the secondary, and then ceases when $e=0$ \citep{murray99}.  

If multiple objects orbit the primary, and the orbiting objects occupy a mean motion resonance, periodic gravitational perturbations will help maintain non-zero eccentricities.  This is the case, for example, for the satellites of Jupiter and Saturn. Jupiter's innermost moon, Io, has a tidal heat flux about $\sim$1--2 W/m$^2$ \citep{spencer00,veeder04}.  The heat flux is so large that the moon has become volcanically active, erupting silicate magmas and sulfur-rich compounds onto its surface \citep{mcewen04}.  Tidal heating can also be an important heat source in bodies with a significant surface layer of H$_2$O (whether it be solid ice or liquid water).  Saturn's icy moon Enceladus, which has a mean radius of 252 km \citep{porco06} and is thus only asteroid-sized, has plumes of water-rich material erupting from tidally induced cracks at its south pole \citep{porco06}, and a localized heat flow between 0.1 and 0.25 W/m$^2$, roughly three times that of the Earth \citep{spencer09}.  Tidal deformation and the geodynamical consequences of tidal heating have also created the bizarre geology on the surfaces of Jupiter's moons Europa \citep{greeley04} and Ganymede \citep{pappalardo04}.

The TRAPPIST-1 planets occupy a mean motion resonance chain with orbital period ratios slightly larger than 24 : 15 : 9 : 6 : 4 : 3 : 2 \citep{unterborn17, luger17}. The resonances maintain the orbital eccentricities of the planets resulting in long-term tidal heating \citep{luger17}.  \citet{luger17} find that assuming Earth-like tidal dissipation rates, TRAPPIST-1b may have a tidal heat flux larger than Io ($\sim 3 \, \mathrm{W/m^2}$) with peaks larger than $10 \, \mathrm{W/m^2}$ due to the periodic change of its orbital eccentricity.  The other six planets in the system would experience weaker tidal heating, with heat fluxes ranging between values similar to the Earth and Io for planets \textit{c--e}, and less than Earth for planets \textit{f--h}. 

Here, we constructed simple compositional and thermal models of the interiors of each of the TRAPPIST-1 planets.  Assuming that the planets are composed of uniform-density (incompressible) rock, iron, and H$_2$O, we explored the range of possible compositional and interior models permitted by the error bars in mass and radius for each planet.  Because the error bars on mass and radius are so large, more sophisticated models including, for example, compression and thermal evolution, are unlikely to give robust and meaningful results.  Our calculations allowed us to determine the precision with which the mass of each planet must be measured to determine whether a given planet harbors a significant reservoir of H$_2$O.  We computed the amount of tidal dissipation in each object based on a uniform viscosity and rigidity, which depend on the volume fractions of ice and rock in each planet. We also determined the expected temperature in the rock mantles of planets b, c, d, e, and f by setting the tidal heat flux equal to the amount of heat that could be removed by solid-state convection.  Some of the planets have globally averaged heat fluxes greater than the heat flux that can trigger a runaway greenhouse effect, which causes irreversible water evaporation and loss \citep{kasting93}. 



\section{Interior structures} \label{section:interior}

\subsection{Prior work} \label{section:interior-prior}

Table \ref{table:masses} summarizes the values of mass, radius, mean density, effective temperature, and orbital period for each of the seven TRAPPIST-1 planets.  The vast majority of the values originate from \citet{wang17}.  The masses range from almost 0.1 $M_{\oplus}$ to 1.63 $M_{\oplus}$, where $M_{\oplus}=5.98\times10^{24}$ kg is the mass of the Earth.  The radii are generally close to the radius of the Earth ($R_{\oplus}=6371$ km).  

Several studies have constrained interior states for the TRAPPIST-1 planets using PREM (Preliminary Earth Reference Model), a density and seismic velocity profile for the Earth \citep{dziewonski81}.  \citet{zeng16} use an equation of state for silicate material based on PREM to constrain iron contents and core sizes in rocky exoplanets as a function of their size and temperature.  These results have been used by \citet{gillon17} and \citet{wang17} to constrain the rock/metal fractions in the TRAPPIST-1 planets.  \citet{wang17} combine the original discovery data for the TRAPPIST-1 system with 73.6 days of K2 observations to obtain updated masses and radii for the planets, which serve as the primary source of mass and radius information for our study.  \citet{wang17} also use equations of state for rock and metal based on PREM to construct interior models of the planets, demonstrating that several of the bodies have densities low enough to contain a significant amount of water in their interiors.  

Another method is to use equations of state for silicate minerals (e.g., MgSiO$_3$) to calculate density and pressure as a function of depth, constrained by the mass and radius of the planet \citep{unterborn17}.  This detailed analysis has been performed for TRAPPIST-1b and 1c.  In these models, phase transitions from one polymorph of MgSiO$_3$ to another are considered, resulting in a mantle which contains layers of enstatite, pyroxene, and garnet, before a final transition to Bridgmanite at a depth of $\sim 1000$ km.  Each of these phases have much lower densities than compressed Bridgmanite, leading to a low-density upper mantle, and permitting a relatively thin H$_2$O layer.

\subsection{Methods} \label{section:interior-methods}
Here, we have calculated the simplest possible interior structures for each of the TRAPPIST-1 planets assuming that the rock, metal, and ice layers in their interiors have uniform densities.  We calculated the suite of structures that fit the mean mass and radius, and explore the range of possible interior structures based on the uncertainties of planetary masses and radii.  Given the large error bars on the masses and radii of the planets, some of which permit planetary mass of zero, simple uniform-density models provide interior structures that are just as valid as those taking into account the pressure and temperature changes with depth inside the bodies \citep[see e.g.,][]{unterborn17}.  

Table \ref{table:masses} shows the mean densities for the planets that we calculated based on their maximum and minimum permitted masses and radii; the large error bars on these quantities yield large error bars on the mean densities, as well. Despite the uncertainties, the densities of the TRAPPIST-1 planets show quite a bit of variation, ranging from values close to the density of water ice under compression, to densities hinting at rocky, and possibly iron-rich compositions. 

 \begin{table*}
 \centering
 \caption{Input parameters for our calculations. Mass, radius, effective temperature, orbital period and eccentricity data are from \cite{wang17}. We have calculated mean densities and uncertainties on density based on the best-fit values and extrema of masses and radii.  \label{table:masses}}
 \vspace{-5mm}
 \begin{tabular}{lllllll}
 \hline
 Planet & Mass (M$_{\oplus})$ & Radius $(R_{\oplus})$ & $\bar{\rho}$ (kg/m$^{3}$) & T$_{eff}$ (K) & Orbital Period (days) & Eccentricity \\
 \hline
 b & $0.79 \pm 0.27$ & $1.086 \pm 0.035$ & $3405^{+1636}_{-1367}$ & 400 & $1.5108739 \pm 0.0000075$ & $0.019 \pm 0.008$\\
 c & $1.63 \pm 0.63$ & $1.056 \pm 0.035$ & $7642^{+4081}_{-3391}$ & 342 & $2.421818 \pm 0.000015$ & $0.014 \pm 0.005$\\
 d & $0.33 \pm 0.15$ & $0.772 \pm 0.030$ & $3960^{+2527}_{-2034}$ & 288 & $4.04982 \pm 0.00017$ & $0.003^{+0.004}_{-0.003}$\\
 e & $^{(\mathrm{a})}0.24^{+0.56}_{-0.034}$ & $0.918 \pm 0.039$ & $^{(\mathrm{a})}1713^{+4790}_{-413}$ & 251 & $6.099570 \pm 0.000091$ & $0.007 \pm 0.003$\\
 f & $^{(\mathrm{a})}0.36^{+0.12}_{-0.061}$ & $1.045 \pm 0.038$ & $^{(\mathrm{a})}1742^{+853}_{-442}$ & 219 & $9.20648 \pm 0.00053$ & $0.011 \pm 0.003$\\
 g & $0.566 \pm 0.036$ & $1.127 \pm 0.041$ & $2183^{+420}_{-354}$ & 199 & $12.35281 \pm 0.00044$& $0.003 \pm 0.002$\\
 h & $^{(\mathrm{a})}0.086^{+0.084}_{-0.017}$ & $0.7150 \pm 0.047$ & $^{(\mathrm{b})}1299^{+1850}_{-19}$ & 167& $18.76626 \pm 0.00068$ & $0.086 \pm 0.032$\\
 \hline
 \end{tabular}
 \floatfoot{Notes: (a) Minimum values of mass and density reported here permit the presence of a small, low-density rock core, which is plausible based on geochemical arguments. (b) We set the mean and the minimum density values to correspond to an ice planet with a small, low-density rock core.}
 \end{table*} 

Our choices of densities for rock, iron, and ice include the effects of compression and phase changes that will occur at depth inside Earth-sized planets.  For iron, we assumed a constant density of $\rho_\mathrm{Fe}=12,000$ kg/m$^3$, consistent with the density of Earth's inner core \citep{dziewonski81, unterborn17}.  For rock in planets b, c, d, and g we assumed a constant density of $\rho_\mathrm{r}=5000$ kg/m$^3$, appropriate for compressed Bridgmanite (MgSiO$_3$) \citep{mccammon16, unterborn17}.  For H$_2$O at low pressures, we assumed a density of $\rho_\mathrm{I}=1000$ kg/m$^3$, which accounts for the possibility of both a solid ice I shell and a liquid water ocean.  For H$_2$O at high pressures, above 209 MPa, we assumed a density of $\rho_\mathrm{hpp}=1300$ kg/m$^3$, which accounts for the possible presence of high-pressure ice polymorphs (hpp) II through VII \citep{hobbs74}, each of which could be present in distinct layers within the planet. When high-pressure ice phases are present, it is likely that ices VI and VII will dominate because they are stable over a broad range of temperatures at the $\sim$tens of GPa central pressures appropriate for TRAPPIST-1 bodies. 

For planets e, f, and h, the minimum permitted planetary densities based on the masses and radii from \citet{wang17} are smaller than the density of high-pressure H$_2$O polymorphs.  In the context of our assumptions, a planet with $\bar{\rho}=\rho_\mathrm{hpp}$ would correspond to a planet composed of pure H$_2$O, which we consider unlikely based on planet and satellite formation models (e.g., \citealt{cw02, morby12, johansen2015}). In Table \ref{table:masses}, we reported a minimum mass corresponding to the mass of a planet with the maximum allowed radius, and $\bar{\rho}\approx\rho_\mathrm{hpp}$.

When determining the interior structures in the low-density limit for planets e and f, and for all of the interior structures for planet h, we adopted a density for rock $\rho_\mathrm{r}=3300$ kg/m$^3$, the density of Prinn-Fegley rock.  Prinn-Fegley rock has been used as a representative composition for the rocky component of the outer planet satellites \citep{mueller88, barr-canup2008}.  This could represent a more hydrated, but compressed rock core. \label{sec:PFrock}

With the exception of TRAPPIST-1c, all of the seven planets have densities low enough to indicate the presence of H$_2$O \citep{wang17,unterborn17}. For these bodies, we began by estimating the maximum depth of the ice I/liquid water layer: $z_{I}=P_I/(\rho_\mathrm{I} g)$, where $g=GM/R_m^2$ is the surface gravity, and $P_I=209$ MPa is the maximum depth at which ice I is possible \citep{hobbs74}.  Subtracting the mass of ice I, we determined the mean density of the remaining planetary material, which allowed us to constrain the relative sizes of the mantles of high pressure ice polymorphs and rock, and the radius of the iron core. Interior models applied by \citet{wang17} suggest that TRAPPIST-1b may contain approximately 25\% water by mass (with a large uncertainty between about 0 and 60\%). Using this constraint of 25\% mass fraction, we were able to obtain a unique interior structure for TRAPPIST-1b that satisfied the mass and radius constraints.

\section{Thermal model} \label{tidalmodel}
Here, we have computed the heat production by tides and heat lost by conduction and convection to constrain tidal heating rates and rock mantle temperatures in each of TRAPPIST-1 exoplanets.  The most common method of estimating tidal heating is to assume that the interior of the secondary object (in this case, the planet) has a Maxwell viscoelastic rheology \citep{love1906, peale78}, characterized by a shear modulus, $\mu$, and viscosity, $\eta$.  More complex rheologies (for example, Andrade or a standard linear solid) can also be used, and yield slightly different behaviors \citep[e.g.,][]{efroimsky12}.  However, given the large uncertainties in the planets' compositions, we chose the Maxwell model because it is a simple model with simple behavior and relatively few parameters.  

In a Maxwell viscoelastic material, heating is maximized in material whose Maxwell time, $\tau=\eta/\mu$ is comparable to the orbital period of the secondary, $P$ \citep{love1906}.  For ice I, the low-density phase of H$_2$O that floats on liquid water, the Maxwell time is comparable to the orbital period of Europa, 85 hours, if the ice is close to its melting point \citep{barr09}.

In a tidally heated body, the interior temperature depends upon the balance and feedback between internal heat generation by, for example, radiogenic and tidal sources, and heat transported by conduction and possibly solid-state convection \citep{schubert01, tackley01, hussmann02, moore06}.  A body whose interior, or portion of its interior, has a viscosity and rigidity close to the values for which $P \sim \tau$, will undergo intense tidal heating.  A body with a slightly higher viscosity, that is, one that is cooler, can be warmed by tidal heating until $P \sim \tau$.  The body can experience inner melting, which in silicate bodies leads to volcanism \citep{peale03}, and in ice/rock bodies, leads to the maintenance of liquid water oceans \citep{ojakangas89}.   Thus, the tidal heating rate in a solid body and its interior structure (namely, composition, temperature, and phase as a function of depth), depend on one another, and the body may reach an equilibrium structure where heat generation is equal to heat loss.

\begin{figure}[ht!]
\centering 
\includegraphics[width=22pc]{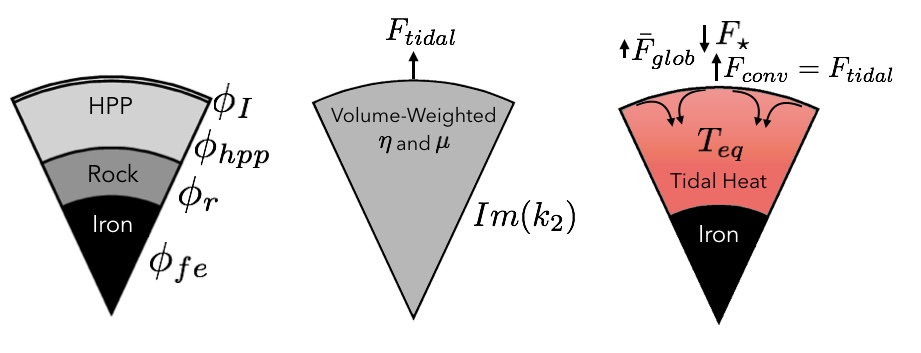}
\caption{Schematic representation of our tidal and thermal model.  \textit{Left:}~internal structure for the planet, composed of water ice I (white), high-pressure ice polymorphs (HPP, gray), rock (dark gray), and iron (black).  The volume fractions of each constituent material were calculated based on the mean density of each planet (see section \ref{section:interior}).  \textit{Middle:}~to calculate the response of each planet to tidal forcing, we constructed a volume-weighted average viscosity and rigidity for the planet based on the $\phi$ values and the rheology of each constituent material.  A value of the imaginary part of the $k_2$ Love number, and the tidal heat flux $F_\mathrm{tidal}$ were calculated.  \textit{Right:}~the temperature in the planet's interior $(T_\mathrm{eq})$ was governed by a balance between heat generation by tides and transport by solid-state convection; the efficiency of both processes depends sensitively on the viscosity and rigidity of the planet.  The net global heat flux $\bar{F}_\mathrm{glob}$ was calculated from the tidal heat flux and the incident sunlight from the star, $F_{\star}$.  \label{fig:schematic}}
\end{figure}

Figure \ref{fig:schematic} shows a schematic illustration of the methods we used to calculate the internal temperature inside each of the TRAPPIST-1 planets based on their different interior structures.  Given values for the volume fractions of each of the planets' constituents ($\phi$), we constructed a volume-averaged viscosity ($\eta$) and rigidity ($\mu$).  Based on these average values, we calculated the response of the planet to the tidal forcing, expressed by the imaginary part of the $k_2$ Love number, $Im(k_2)$.  This value was used to calculate the surface heat flux from tidal dissipation $F_\mathrm{tidal}$. The heat generated from tidal dissipation is likely removed from the planets' deep interiors via solid-state convection, the dominant heat transfer mechanism in the terrestrial planets and ice/rock satellites in our Solar System.  To determine the temperature in the planets' deep interiors, we balanced $F_\mathrm{tidal}$ and the convective heat flux ($F_\mathrm{conv}$), and found the value of temperature ($T_\mathrm{eq}$) for which $F_\mathrm{conv}=F_\mathrm{tidal}$.  If $T_\mathrm{eq}$ is above the solidus temperature for rock, the planet may possess a magma ocean in its interior.  The globally averaged energy flux on the planet is the sum of the energy flux from the mean stellar irradiation ($F_\star$) and tidal heating,
\begin{equation}
   \label{eq:Fglob}
      \overline{F}_\mathrm{glob} = F_\star + F_\mathrm{tidal} \, ,
\end{equation}
where $F_{\star}$ is taken at the top of the planet's atmosphere.  The stellar radiation can be expressed as
\begin{equation}
   \label{eq:Fstar}
      F_\star = \frac { L_\star \left( 1 - A_\mathrm{B} \right) } { 16 \pi a^2 \sqrt{1 - e^2} } \, ,
\end{equation}
where $L_\star$ is the stellar luminosity, $A_\mathrm{B}$ is the Bond albedo of the planet, $a$ and $e$ are the semi-major axis and the eccentricity of the planet's orbit, respectively. Greenhouse and cloud feedbacks were neglected in the calculation of $F_{\star}$. The albedo of each of the planets was set to 0.3, an Earth-like value.

\subsection{Heat generation}
Here, we have focused on tidal heating as the dominant source of energy for heating the interiors of the TRAPPIST-1 planets.  This is because the magnitude of tidal heating expected in rock- or ice-dominated bodies at the orbital periods associated with the TRAPPIST-1 planets, a few to ten watts per meter squared, is several orders of magnitude higher than the heat flows associated with radiogenic heating, which is typically of the order of tens of milliwatts per meter squared \citep{henning09, henning14}.  Residual heat of accretion is also a possible heat source in any solid planet.  The energy per unit mass released during accretion, $E_\mathrm{acc} \sim (3/5)GM/R$ is sufficient to melt planets as large as the TRAPPIST-1 bodies, and is still a significant contribution to the Earth's heat flux.  Even if the TRAPPIST-1 bodies still retained 20\% of their accretional heat, the resulting heat flux would still be much smaller than the tidal heat flux.

The tidal heat flux was calculated using a viscoelastic model for a homogeneous body described by \citet{dobos15}, which was originally developed by \citet{henning09}. Although the planet was assumed to be homogeneous, we mimicked the effect of multiple materials by determining the effective viscosity and rigidity for each layers and weighting them by their volume fractions. To approximate the material properties of a planet composed of several different materials, we calculated the volume fraction of each material contained in the planet: for a given material, $i$, the volume fraction $\phi=V_i/V_\mathrm{tot}$, where $V_\mathrm{tot}=(4/3)\pi R_{pl}^3$ is the total volume of the planet.  We calculated values of the volume fractions of ice I ($\phi_I$), high-pressure ice polymorphs ($\phi_\mathrm{hpp}$), rock ($\phi_\mathrm{r}$), and iron ($\phi_\mathrm{Fe}$) contained in each planet, using the mass and radius of the planet as constraints, assuming uniform densities for each of the materials. A single uniform viscosity and rigidity for the planet was approximated by
\begin{equation}
\label{eq:composite}
\eta \approx \phi_\mathrm{I}\eta_\mathrm{I} + \phi_\mathrm{hpp} \eta_\mathrm{hpp} + \phi_\mathrm{r} \eta_\mathrm{r} + \phi_\mathrm{Fe} \eta_\mathrm{Fe}, 
\end{equation}
where $\eta$ is the viscosity of the material.  We used a similar relationship to construct a single value of the shear modulus ($\mu$) that approximates the behavior of the entire planet:
\begin{equation}
\label{eq:mu-composite}
\mu \approx \phi_\mathrm{I}\mu_\mathrm{I} + \phi_\mathrm{hpp} \mu_\mathrm{hpp} + \phi_\mathrm{r} \mu_\mathrm{r} + \phi_\mathrm{Fe} \mu_\mathrm{Fe}. 
\end{equation} 

The total amount of energy produced by tidal dissipation in a synchronously rotating body \citep{segatz88},
\begin{equation}
   \label{viscel}
       \dot{E}_\mathrm{tidal} = - \frac {21} {2} Im(k_2) \frac {R^5 \omega^5 e^2} {G} \, ,
\end{equation}
where $\omega=2 \pi/P$ is the orbital frequency of the planet, $P$ is the planet's orbital period, $G$ is the gravitational constant, $e$ is the planet's eccentricity, and $R$ is its radius.  The quantity $Im(k_2)$ is the imaginary part of the $k_2$ Love number, which describes how the planet's gravitational potential is disturbed by its tidal distortion.  The value of $Im(k_2)$ depends on the structure and rheology in the body \citep{segatz88}. 

There are a variety of different ways of calculating $Im(k_2)$ for a solid planet. In some studies, $Im(k_2)$ is expressed as $k_2/Q$, where $Q$ is a constant parameter describing the fraction of orbital energy dissipated per tidal cycle (see, e.g., \citealt{murray99} for discussion, and \citealt{papaloizou17} for example application of this approach to the TRAPPIST-1 system).  The value of $Q$ for solid planetary bodies is not well-constrained, and is generally thought to be between $Q\sim10$ and 200 \citep{goldreich66, murray99}.

A constant-$Q$ approach, however, does not take into account the feedback between the thermal state of a planet and its response to tidal forces.  A warm planet will deform more in response to tidal forces than a cold, rigid planet. Thus, it is more accurate to calculate $Im(k_2)$ based on the viscosity and rigidity of the planet's interior.  A common approach is to assume that the planet is composed of a single material (for example, rock).  Given the density, shear modulus, and viscosity of the planetary material, and an assumption about the rheology (for example Maxwell, Andrade, etc., \citet{efroimsky12}), it is possible to derive an analytic expression for $Im(k_2)$ (e.g., \citealt{harrison63,henning09,dobos15}, see Eq. \ref{Imk2} here).

A more sophisticated approach is to assume that the planet is composed of, perhaps two or three materials, each of which has a different density, shear modulus, and viscosity. In this case, analytic, although quite cumbersome expressions can still be obtained for $Im(k_2)$ \citep{harrison63,remus12,beuthe13,remus2015}.  Many of these studies rely upon simplifying assumptions to obtain closed-form analytic solutions (e.g., \citealt{remus12,remus2015}); in most cases, these simplifying assumptions may not apply to the TRAPPIST-1 planets.  Another approach is to calculate $Im(k_2)$ numerically, by representing the planet by ten or more layers, each of which has a density, shear modulus, and viscosity dictated by the composition of the layer, but also thermal state of the material (e.g., \citealt{sabadini82,moore06,roberts08}).

Given the large uncertainties in the masses and radii of the TRAPPIST-1 planets, calculation of interior thermal states using more sophisticated models (for example, one in which the heat generation and transport in each different compositional layer inside the planet is treated separately) would be premature. Therefore, we used a Maxwell viscoelastic rheology to calculate the value of \textit{Im}($k_2$):
\begin{equation}
   \label{Imk2} 
       - Im(k_2) = \frac {57 \eta \omega} { 4 \bar{\rho} g R\left[ 1 + \left( 1 + \dfrac { 19 \mu } { 2 \bar{\rho} g R } \right)^2 \dfrac { \eta^2 \omega^2 } { \mu^2 } \right] } \, ,
\end{equation}
where $g=GM/R^2$ is the planet's surface gravity, $\eta$ is the viscosity (equation \ref{eq:composite}) and $\mu$ (equation \ref{eq:mu-composite}) is the shear modulus of the planet \citep{henning09}. 

The globally averaged tidal heat flux was calculated from
\begin{equation}
F_\mathrm{tidal}=\frac{\dot{E}_\mathrm{tidal}}{4 \pi R^2}. \label{eq:Ftidal}
\end{equation}

It is worth noting that this tidal heating model is valid only for small orbital eccentricities, that is, for $e{\lesssim}0.1$. At higher eccentricities, additional higher-order terms must be included in the tidal potential, which can increase the amount of heating \citep{MignardII}.  Fortunately, this constraint is fulfilled for each of the seven planets, according to the data of \citet{gillon17} and \citet{wang17}.  In our model, we did not consider gravitational interactions of the planets with each other, only the star-planet (two-body) tidal interaction was considered in each case.

\subsection{Heat loss}  
The tidal heat generated inside the planet was assumed to be transported to the surface by solid-state convection, then conducted across a surface boundary layer where it is radiated into space. The heat flux across a convecting mantle,
\begin{equation}
F_\mathrm{conv}= \frac{k_\mathrm{therm} \Delta T}{D} \mathrm{Nu}, \label{eq:cond}
\end{equation}
where $k_{therm}$ is the thermal conductivity, $\Delta T=T_\mathrm{mantle}-T_\mathrm{surf}$ is the difference in temperature between the hot, convecting mantle ($T_\mathrm{mantle}$) and $T_\mathrm{surf}$ is the temperature at the surface, and $D$ is the thickness of the convecting layer.  The parameter $\mathrm{Nu}$, the Nusselt number, describes the relative efficiency of convective heat transport versus conductive heat transport.  In an internally heated planet with a Newtonian rheology, $\mathrm{Nu}$ is related to the thermal, physical, and rheological properties of the mantle \citep{SM2000},
\begin{equation}
\mathrm{Nu}=0.53 \theta^{-4/3}\mathrm{Ra}^{1/3}, \label{eq:Nudef}
\end{equation}
where $\theta=(Q^*\Delta T/R_\mathrm{G}T_\mathrm{mantle}^2)$, $Q^*$ is the activation energy in the flow law for the convecting material and $R_\mathrm{G}=8.314$ J/mol/K is the universal gas constant.  The parameter $\mathrm{Ra}$ is the Rayleigh number, which describes the vigor of convection,
\begin{equation}
Ra=\frac{\rho g \alpha \Delta T D^3}{\kappa \eta(T_\mathrm{mantle})}, \label{eq:Radef}
\end{equation}
where the coefficient of thermal expansion $\alpha$, thermal diffusivity $\kappa_\mathrm{therm}=k_\mathrm{therm}/(\rho C_p)$, $C_p$, and $\eta(T_\mathrm{mantle})$ is the viscosity evaluated at the mantle temperature.  Using Eq. \ref{eq:Radef} to evaluate the Rayleigh number in Eq. \ref{eq:Nudef}, we obtained an expression for the convective heat flux which did not depend on the thickness of the convecting layer, or the temperature at the surface of the planet \citep{SM2000,barr08},
\begin{equation}
F_\mathrm{conv}=0.53 \bigg(\frac{Q^*}{R_G T_\mathrm{mantle}^2}\bigg)^{-4/3} \bigg(\frac{\rho g \alpha k_\mathrm{therm}^3}{\kappa_\mathrm{therm} \eta(T_\mathrm{mantle})}\bigg)^{1/3}. \label{eq:heatflux}
\end{equation}
We assumed that the majority of tidal heat will be produced in, and transported within, the rock mantle of the planet.  Thus, the parameters in Eq. \ref{eq:heatflux} were evaluated using the properties of rock: $Q^*=333$ kJ/mol (see Sect. \ref{sec:material-properties-rock}), $k_\mathrm{therm}=3.2$ W/m/K, $\alpha=3 \times 10^{-5}$ 1/K, and $C_p=1200$ J/kg \citep{SM2000}.

\begin{figure}[ht!]
\centering 
\includegraphics[width=22pc]{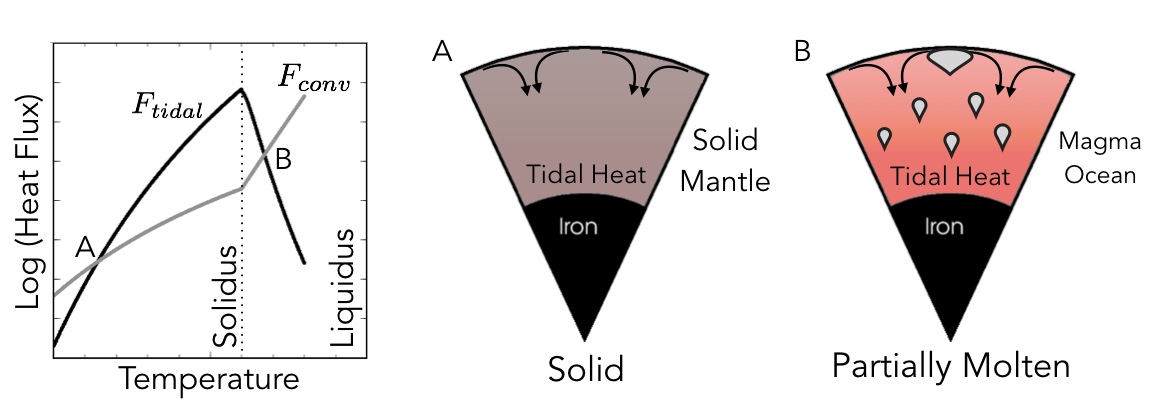}
\caption{ \textit{Left:}~schematic representation of the heat flux (in log units) from tidal heating (black line) and convective heat transport (gray) as a function of temperature.  Two equilibria are possible: point (A) represents a balance between tidal heating and convection in a purely solid mantle ($T_\mathrm{eq}$ is less than the solidus temperature $T_\mathrm{s}$).  This state is depicted schematically in the middle panel of the figure.  Point (B) represents an equilibrium for which $T_\mathrm{eq}>T_\mathrm{s}$, implying a partially molten rock mantle exists within the planet.  This state is shown schematically in the right panel of the figure. \label{fig:equilibria}}
\end{figure}

To determine the equilibrium (or equilibria) between $F_\mathrm{conv}$ and $F_\mathrm{tidal}$, we evaluated Eq. \ref{eq:Ftidal} and Eq. \ref{eq:heatflux} for a variety of values of $T_\mathrm{mantle}$, the temperature in the rock mantle of each planet.  Two equilibria are possible between the tidal heat flux and convective cooling flux which correspond to two equilibrium temperatures of $T_\mathrm{mantle}$, which we referred to as $T_{eq}$ \citep{moore06, dobos15}.  The left panel of Fig. \ref{fig:equilibria} illustrates schematically how the tidal and convective heat fluxes vary as a function of temperature.  One equilibrium can exist at a temperature well below the solidus, corresponding to a balance between tidal heat generation and transport in a purely solid planetary mantle (point (A) in the left panel, and also see the middle panel of Fig. \ref{fig:equilibria}).  In this case, the amount of tidal heat generated is sufficiently low to be removed by convection, even if the convective flow is relatively sluggish. This equilibrium is unstable -- if the tidal heat generation increases, convection cannot remove the additional tidal heat \citep{moore03, dobos15}.  Depending on the tidal forcing frequency and the rheology of the planet, a second equilibrium can be achieved in which the planet is partially molten (point (B) in the left panel, and the right panel of Fig. \ref{fig:equilibria}).  Tidal heat generation decreases as a function of temperature and melt fraction above the solidus because partially molten rock is less dissipative than warm, solid rock.  However, the convective heat flux increases sharply as the presence of melt decreases the mantle viscosity. This high-temperature equilibrium is stable \citep{moore03, dobos15} -- as the tidal heating rate increases, the viscosity of the mantle decreases, permitting more efficient convective heat flow and resulting in cooling of the mantle.

\subsection{Runaway greenhouse limit}

To characterize the habitability of the planets, we determined whether the planet might enter a runaway greenhouse state. When the globally averaged heat flux on a planet exceeds a critical value, then surface water slowly evaporates. The water molecules in the upper atmosphere disintegrate to hydrogen and oxygen due to photodissociation by ultraviolet photons, and then hydrogen molecules can easily escape to space. Hence, entering the runaway greenhouse phase could imply that water molecules would not be able to re-form later, even if the temperature decreases \citep{kasting93}.

To determine the runaway greenhouse limit, we used the formulation of \citet[Chapter 4]{pierrehumbert10}. If the globally averaged energy flux on the planet from stellar irradiation and tidal heat ($\bar{F}_{glob}$) exceeded the minimum energy flux on top of a water-rich atmosphere then it triggers the runaway greenhouse effect:
\begin{equation}
F_\mathrm{RG} = o \frac{ \sigma (l / R_\mathrm{water})^4 } { \ln \left( p^\star / \sqrt{ 2 p_0 g / \kappa_0} \right)^4 }  \, , \label{eq:Frg}
\end{equation}
where $o=0.7344$, $\sigma$ is the Stefan-Boltzmann constant, $l=2.425 \cdot 10^6$~J/kg is the latent heat, $R_\mathrm{water}=461.5$~J/(kg K) is the gas constant for water vapor, $\kappa_0=0.055$ is the gray absorption coefficient, $p_0=10000$~Pa is the reference pressure for absorption, $p^\star=p_\mathrm{ref} \exp \left( \frac{l} {R T_\mathrm{ref} } \right)$, $p_\mathrm{ref}=610.616$~Pa and $T_\mathrm{ref}=273.13$~K.

\subsection{Viscoelastic material properties} \label{sec:material-properties}

\subsubsection{Rock} \label{sec:material-properties-rock}

For rock, we used the viscosity and rigidity model used for exomoon interiors described in \citet{dobos15} which is based on the works of \citet{henning09}, \citet{moore03}, and \citet{fischer90}.  The behavior of the rock changes as a function of temperature, as the temperature crosses the solidus point ($T_\mathrm{s}=1600$ K), across a ``breakdown'' point ($T_\mathrm{b}=1800$ K) at which the volume fraction of solid rock crystals is equal to the volume fraction of melted rock.  At this point, the behavior of the crystal and melt mixture begins to behave more like a liquid \citep{renner2001}.  When the rock is completely melted, (above the liquidus ($T>T_l$, where $T_\mathrm{l}=2000$ K) its viscosity and shear modulus are quite small.  For temperatures below the solidus, $\mu_{r}=50$ GPa, and the viscosity is described by,
\begin{equation}
\eta_\mathrm{r}(T)=\eta_{0}\exp\bigg(\frac{Q^*_{r}}{R_GT}\bigg),
\end{equation}
where $\eta_0=2.13 \times 10^6$ Pa s \citep{henning09}, the activation energy $Q^*_{r}=333$ kJ/mol \citep{fischer90}, and $R_G=8.314$ J/(mol~K) is the gas constant. 

For temperatures above the solidus, but below the breakdown temperature ($T_s < T < T_b$), the shear modulus depends on temperature \citep{fischer90},
\begin{equation}
\mu_\mathrm{r}(T) = 10^{\left(\frac{\mu_1}{T}+\mu_2\right)},
\end{equation}
where $\mu_1=8.2\times 10^{4}$ K and $\mu_2=-40.6$.  In this regime, the viscosity of rock depends on the volume fraction of melt present \citep{moore03}, $f$, which varies linearly with temperature between the liquidus and solidus,
\begin{equation}
\eta_\mathrm{r}(T)=\eta_0\exp\bigg(\frac{E}{R_\mathrm{G}T}\bigg) \exp(-Bf),
\end{equation}
where $B$ is the melt fraction coefficient \citep[$10 \leq B \leq 40$,][]{moore03}. Above the breakdown temperature, but still below the liquidus, the shear modulus, now controlled by the behavior of the volumetrically dominant silicate melt, is a constant at $\mu_\mathrm{r}=10^{-7}$ Pa.  The viscosity follows the Einstein-Roscoe relationship \citep{moore03},
\begin{equation}
\eta_\mathrm{r}(T)=10^{-7}\textrm{ Pa s} \exp\bigg(\frac{40,000\textrm{ K}}{T}\bigg)(1.35 f - 0.35)^{-5/2}. \label{eq:rock-er}
\end{equation}
Above the liquidus, $\mu_\mathrm{r}=10^{-7}$ Pa s, and the viscosity continues to depend on temperature, obeying Eq. (\ref{eq:rock-er}) with $f=0$.

\subsubsection{Ice I and liquid water}

For ice I, deformation was assumed to occur by volume diffusion \citep{GoldsbyKohlstedt2001,barr08}, for which the viscosity depends on temperature and grain size $d$,
\begin{equation}
\eta_\mathrm{I}(T)=\frac{R_\mathrm{G}Td^2}{14V_\mathrm{m}D_{ov}}\exp\bigg(\frac{Q^*_\mathrm{ice}}{R_\mathrm{G}T}\bigg),
\end{equation}
where the molar volume $V_\mathrm{m}=1.95\times 10^{5}$ m$^3$/mol,  diffusion coefficient $D_{ov}=9.10\times 10^{-4}$, and activation energy $Q^*_\mathrm{ice}=59.4$ kJ/mol.  Grain sizes for ice in natural systems on Earth are limited to a few millimeters, due to the presence of silicate particles which inhibit grain growth \citep{durand2006}.  Temperature and strain rate conditions in terrestrial glaciers are somewhat similar to those expected in the outer planet satellites in our Solar System, leading to the suggestion that $d \lesssim 1$ mm in the ice I mantles of those bodies \citep{barr07}.  In the absence of other information, it is reasonable to assume that the process would work in a similar fashion on other planets: we adopted $d=1$ mm as a nominal value.  Similar to rock, the shear modulus for ice I is weakly temperature-dependent \citep{gammon83},
\begin{equation}
\mu_\mathrm{I}(T)=\mu_{I,T_\mathrm{m}}\bigg[\frac{1-aT}{1-aT_\mathrm{m}}\bigg],
\end{equation}
where $\mu_{I,T_\mathrm{m}}=3.39\times 10^9$ Pa, $a=1.418\times 10^{-3}$ K$^{-1}$. 

The melting temperature of ice I is $T_\mathrm{m}=273$ K.  Above the melting point, $\mu_\mathrm{I}=0$, and $\eta_\mathrm{I}=10^{-3}$ Pa~s which we used for the liquid water layers.

\subsubsection{High-pressure ice polymorphs (HPPs)}

Experimental data for the ductile behavior of the high-pressure ice polymorphs are more limited than experimental studies for ice I.  It is known that each of the high-pressure polymorphs has a distinct flow law, and many of them are strongly non-Newtonian, so that the viscosity depends strongly on stress \citep{durham97}. Similar behavior is observed in rock far from the melting point (e.g., \citealt{karato95}).  We used a flow law that approximates the laboratory-determined flow behavior for ices VI and VII from \citet{durham97, durham-correction}.  Our tidal model assumed a Maxwell behavior, in which viscosity does not depend on stress, so we assumed a constant stress of 0.3~MPa.  The approximate hpp flow law,
\begin{equation}
\eta_\mathrm{hpp}=A \exp \left(\frac{Q^*_{hpp}}{R_\mathrm{G}T}\right),
\end{equation}
where $A=6.894 \times 10^{-7}$ and $Q^*_{hpp}=110$ kJ/mol, gives a viscosity at the approximate melting point of ices VI and VII, $T_\mathrm{m}=285$ K, of $10^{14}$ Pa s, somewhat similar to nominal near-melting-point viscosities for ice I.  Data for the shear modulus of ices VI and VII are also limited, but at temperatures close to the melting point, $\mu_\mathrm{hpp}\approx 3.5\times 10^{10}$ Pa \citep{shimizu96}, where this value represents an average over the range for which experimental exist, which roughly corresponds to the pressure conditions expected within the TRAPPIST-1 planets' interiors.  We assumed $\mu_\mathrm{hpp}$ was independent of temperature below the melting point.  Above the melting point, high-pressure ices take on the properties of liquid water: $\mu_\mathrm{hpp}=0$, and $\eta_\mathrm{hpp}=10^{-3}$ Pa~s.  

\subsubsection{Iron}

The viscosity and rigidity of iron were assumed to be constant, and equal to the values deduced for the inner core of the Earth. The rigidity of iron, $\mu_\mathrm{Fe}=1.575\times 10^{11}$ Pa, based on PREM \citep{dziewonski81}.  The viscosity of iron in Earth's inner core is thought to lie between $2-7\times 10^{14}$~Pa~s \citep{koot2011}; we used $\eta_\mathrm{Fe}=4.5\times 10^{14}$~Pa~s.  

\subsection{Benchmark}
To test the validity of our approach, we first used our methods to calculate the equilibrium interior temperature and tidal heat flux from Jupiter's moon Io, for which the tidal heat flux has been measured by spacecraft to be $\sim 1$W/m$^2$ \citep{spencer00, veeder04}.  We used the following parameters for Io: mass $M=8.93\times10^{22}$ kg, rock volume fraction $\phi_{r}=0.95$, iron fraction $\phi_{fe}=0.05$ \citep{schubert04}, orbital period $P=1.769$ days, eccentricity $e=0.0041$, and $R=1821$ km.  With these values, our model yielded a tidal heat flux $F_{tidal}=0.38$ W/m$^2$, comparable in magnitude to that observed, and an equilibrium rock mantle temperature $T_{eq}=1671$ K, above the rock solidus, and consistent with the inference of a partially molten mantle beneath Io's surface \citep{khurana2011}.   

\subsection{Uncertainties}
Because the uncertainties in the measured planetary radius and mass are quite large (in some cases, they permit zero mass), it is important to provide limits to our calculations, as well. Here, we have reported a range of interior structures for the lowest and highest densities that can be estimated from the minimum and maximum radii and masses (see Table \ref{table:masses}). In the cases of planets e, f, and h, we have modified some values in order to avoid the possibility of zero mass and the lack of a rocky core. For these planets, when determining the interior structure for the low-density case (and also for the other density values in the case of planet h), we assumed the rock to be Prinn-Fegley rock, as described in Sect. \ref{section:interior-methods}. These calculations with the lowest and highest densities effectively ``bracket'' the range of possible interior structures.

With these new structures, we recalculated the tidal fluxes, as well. Also, the uncertainties of the orbital eccentricities propagate into the tidal heating calculation. For this reason, we recalculated the tidal fluxes using the lowest and highest eccentricity values given in Table \ref{table:masses}. Hence, we got nine different  values for the tidal flux (one mean value for each density and the upper and lower limits provided by the minimum and maximum eccentricities).

When calculating the incident stellar flux, the largest uncertainty comes from the unknown albedo of the planets. For these calculations, we used the albedo value of the Earth, $A_\mathrm{B}=0.3$. Because the true albedos of the planets are not known, we repeated the calculations for 0.1 and 0.5 planetary albedos, too, and used the difference as error bars for the stellar (and consequently for the global) flux of the planets.

For the runaway greenhouse flux, the uncertainty came from the change of the surface gravity which depends on the mass and radius of the planet which are highly uncertain. We used the maximum and minimum values of stellar flux obtained from the minimum and maximum density calculations, and indicate them as uncertainties of the global flux of the planets. The uncertainties of the tidal flux were neglected here, because those are two orders of magnitude lower compared to the uncertainties in the stellar flux.

\section{Results and discussion} \label{results} \label{sec:results}
Table \ref{table:structures} summarizes the volume fraction of each constituent material in each of the TRAPPIST-1 planets. Because of the large uncertainties in the measured mass and radius of the planets, we first calculated three end-member structures, corresponding to the highest-density planet, a structure for the mean-density planet, in which the planet has the best-fit mass and radius from \citet{wang17}, and the low-density planet corresponding to the lowest permitted density.  For planet h, the mean density value was very close to the minimum that could be obtained with assuming a low-density rock core, thus, we reported only the mean and maximum density cases.  

Given the large error bars on mass and radius for each planet, it was not always possible to determine the bulk composition or the number of different compositional layers inside the planet -- for planets b, c, d, and e, water-free structures were permitted if the planets had densities close to their upper limits.  Likewise, each of these planets could be composed of rock and H$_2$O, completely devoid of iron, if their true densities were close to their lower limits.  For planets b, c, d, and e, we also computed the limit where the structure qualitatively changed, for example, one of the layers disappeared or a new one appeared, consistent with each planet's possible values of mass and radius.    

 \begin{table*}
 \centering
 \caption{Inner structure of the planets for maximum, minimum and mean bulk densities: volume fraction ($\phi$) and radius ($R$) of the top of iron, rock, high-pressure ice polymorphs, and ice I/water layers in each of the seven TRAPPIST-1 exoplanets. Maximum mass and minimum radius (as indicated in Table \ref{table:masses}) were assumed for the calculations indicated by $\bar{\rho}_\mathrm{min}$, and similarly, minimum mass and maximum radius were assumed for the $\bar{\rho}_\mathrm{max}$ cases.  \label{table:structures}}
 \vspace{-5mm}
 \begin{tabular}{cccccccccc}
 \hline
 Planet & density & $\phi_\mathrm{Fe}$ & $\phi_\mathrm{r}$ & $\phi_\mathrm{hpp}$ & $\phi_\mathrm{I}$ & $R_{fe}$ [km] & $R_r$ [km] & $R_{hpp}$ [km] & $R_I$ [km] \\
 \hline 
   & $\bar{\rho}_\mathrm{max}$ & 0.006 & 0.994 & --    & --    & 1204 & 6696 & --   & -- \\
 b & $\bar{\rho}$              & 0.121 & 0.221 & 0.644 & 0.014 & 3418 & 4839 & 6887 & 6919 \\
   & $\bar{\rho}_\mathrm{min}$ & --    & 0.201 & 0.777 & 0.021 & --   & 4185 & 7090 & 7142 \\
 \hline 
   & $\bar{\rho}_\mathrm{max}$ & 0.960 & 0.040 & --    & --    & 6418 & 6505 & --   & -- \\
 c & $\bar{\rho}$              & 0.377 & 0.623 & --    & --    & 4862 & 6728 & --   & -- \\
   & $\bar{\rho}_\mathrm{min}$ & --    & 0.799 & 0.191 & 0.011 & --   & 6449 & 6925 & 6951 \\
 \hline 
   & $\bar{\rho}_\mathrm{max}$ & 0.212 & 0.788 & --    & --    & 2820 & 4727 & --   & -- \\
 d & $\bar{\rho}$              & --    & 0.721 & 0.256 & 0.023 & --   & 4410 & 4880 & 4918 \\
   & $\bar{\rho}_\mathrm{min}$ & --    & 0.173 & 0.783 & 0.044 & --   & 2846 & 5034 & 5110 \\
 \hline 
   & $\bar{\rho}_\mathrm{max}$ & 0.215 & 0.785 & --    & --    & 3353 & 5600 & --   & -- \\
 e & $\bar{\rho}$              & --    & 0.115 & 0.848 & 0.038 & --   & 2841 & 5774 & 5849 \\
   & *$\bar{\rho}_\mathrm{min}$ & --    & 0.007 & 0.947 & 0.046 & --   & 1158 & 6003 & 6097 \\
 \hline
   & $\bar{\rho}_\mathrm{max}$ & -- & 0.352 & 0.627 & 0.021 & -- & 4529 & 6371 & 6416 \\
 f & $\bar{\rho}$              & -- & 0.122 & 0.850 & 0.029 & -- & 3299 & 6593 & 6658 \\
   & *$\bar{\rho}_\mathrm{min}$ & -- & 0.005 & 0.959 & 0.036 & -- & 1209 & 6816 & 6900 \\
 \hline
   & $\bar{\rho}_\mathrm{max}$ & -- & 0.354 & 0.628 & 0.018 & -- & 4893 & 6877 & 6919 \\
 g & $\bar{\rho}$              & -- & 0.240 & 0.740 & 0.020 & -- & 4463 & 7132 & 7180 \\
   & $\bar{\rho}_\mathrm{min}$ & -- & 0.145 & 0.833 & 0.022 & -- & 3908 & 7386 & 7441 \\
 \hline
   & *$\bar{\rho}_\mathrm{max}$ & -- & 0.930 & 0.031 & 0.039 & -- & 4154 & 4200 & 4256 \\
 h & *$\bar{\rho}$               & -- & 0.012 & 0.907 & 0.081 & -- & 1031 & 4429 & 4555 \\
 \hline
 \end{tabular}
 \floatfoot{Note: *For these models, $\rho_\mathrm{r}=3300$ kg/m$^3$ (see Sect. \ref{sec:PFrock}).}
 \end{table*}
 
The values in Table \ref{table:structures} served as inputs in our calculations of heat generation by tides and loss by interior processes. Table \ref{table:tidalcalc} summarizes the values of equilibrium mantle temperature, tidal heat flux, and global energy flux at the top of the atmosphere from our geophysical models.  In addition to the uncertainties provided by the densities (mass and radius values), the orbital eccentricity altered the tidal heating flux on the planet, too. Calculating with the three density and three eccentricity values for each planet, we obtained nine different results for the equilibrium temperatures and the tidal heating fluxes. (For planet h it resulted only in six different values.) For simplicity, we only presented the one case calculated with the mean density and mean eccentricity value, and the upper and lower uncertainties from among all other cases.  The equilibrium values of mantle temperature obtained in our calculations, as well as the tidal heat flux, $\bar{F}_{glob}$, and threshold heat flux for a runaway greenhouse, $F_{RG}$, are summarized in Table \ref{table:tidalcalc}.  

 \begin{table}
 \centering
 \caption{Results of the tidal heating calculations: equilibrium temperature in the mantle ($T_\mathrm{eq}$ in Kelvin), tidal heat flux ($F_\mathrm{tidal}$), globally averaged energy flux at the the top of the atmosphere ($\overline{F}_\mathrm{glob}$) and limit for the runaway greenhouse flux ($F_\mathrm{RG}$). All heat fluxes are reported in W/m$^2$.  The missing lower error for planet d indicates that the equilibrium temperature was not found in all investigated cases because of the low tidal heating rate, and as a consequence, there was no value below 1618 K. Similarly, no equilibrium temperature was found for planets g and h. Calculations were made using three different planetary albedos for each planet: 0.1, 0.3 and 0.5. \label{table:tidalcalc}}
 \vspace{-5mm}
 \begin{tabular}{ccccc}
 \hline
Planet & $T_\mathrm{eq}$ & $F_\mathrm{tidal}$ & $\overline{F}_\mathrm{glob}$ & $F_\mathrm{RG}$  \\
 \hline 
 b & $1709^{+28}_{-10}$  & $2.68^{+1.33}_{-1.81}$  & $1014 \pm 289$ & $283^{+11}_{-15}$  \\
 c & $1666^{+33}_{-64}$  & $1.32^{+0.3}_{-0.47}$   & $540 \pm 154$  & $308^{+14}_{-18}$ \\
 d & $1618^{+33}$        & $0.16^{+0.35}_{-0.16}$  & $271 \pm 78$   & $277^{+14}_{-20}$ \\
 e & $1629^{+10}_{-29}$  & $0.18^{+0.09}_{-0.18}$  & $157 \pm 45$   & $258^{+38}_{-7}$ \\
 f & $1621^{+8}_{-15}$   & $0.14^{+0.05}_{-0.14}$  & $91 \pm 26$    & $262^{+10}_{-7}$ \\
 g & --                  & $0 \pm 0$               & $61 \pm 18$    & $270^{+4}_{-4}$  \\
 h & --                  & $0 \pm 0$               & $35 \pm 10$    & $244^{+22}_{-1}$ \\
 \hline
 \end{tabular}
 \end{table}

\subsection{Planet b}

\begin{figure}[ht!]
\centering 
\includegraphics[width=20pc]{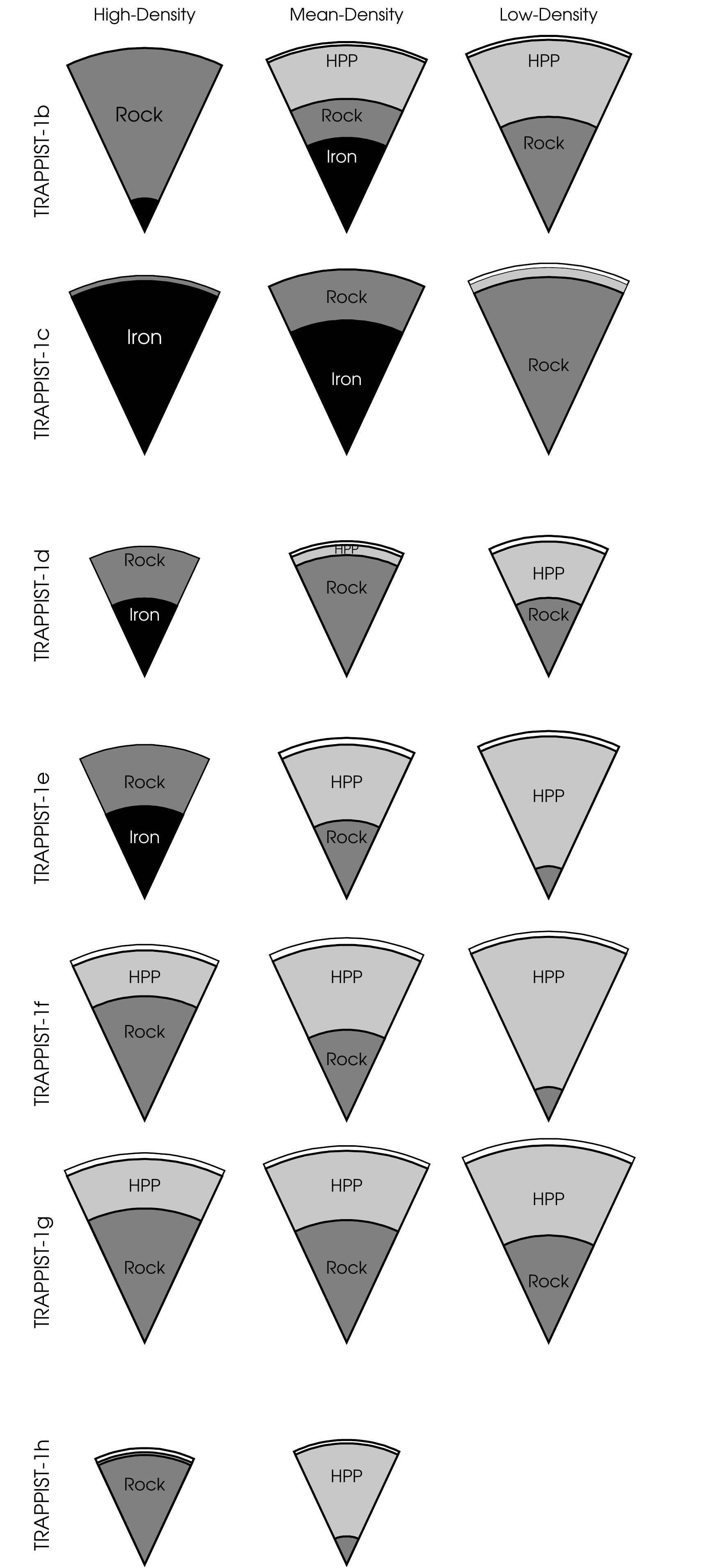}
\caption{Three end-member structures for each of the TRAPPIST-1 planets, assuming constant densities for ice I and liquid water (white) with $\rho_\mathrm{I}=1000$ kg/m$^3$, high-pressure ice polymorphs ($\rho_\mathrm{hpp}=1300$ kg/m$^3$, light gray), rock ($\rho_\mathrm{r}=5000$~kg/m$^3$, dark gray) or low-density rock as indicated in Table \ref{table:structures} ($\rho_\mathrm{r}=3300$~kg/m$^3$, dark gray, see Sect. \ref{sec:PFrock}), and iron ($\rho_\mathrm{Fe}=12000$ kg/m$^3$, black), corresponding to the maximum density permitted by orbital solutions (left), the mean density (middle), and lowest density (right).  The sizes of the planets and thicknesses of ice polymorph, rock, and iron layers are quantitatively accurate; but in some cases, the thickness of the outer water/ice I layer has been expanded slightly to show detail. \label{fig:planets}}
\end{figure}

The top row of Fig. \ref{fig:planets} illustrates our proposed interior structures for TRAPPIST-1b.  The mass of planet b is comparable to the planet Venus, but with a radius of $1.086 \pm 0.035R_{\oplus}$, its mean density of $3405$ kg/m$^{3}$ is significantly lower than the mean densities of either Venus or Earth. Given our assumed densities for rock and iron, the highest permitted density for TRAPPIST-1b, $\rho_{max}=5041$ kg/m$^3$, permitted the planet to be composed of these two materials, without a significant reservoir of H$_2$O (see left panel of Fig. \ref{fig:planets}).  However, this represents an extreme end-member scenario.  The absolute lower limit of mean density, $\rho_{min}=2038$ kg/m$^3$, permitted an interior structure with no iron, solely composed of rock and H$_2$O (see right panel of Fig. \ref{fig:planets}).  Given the current estimates of mass and density, the true structure of TRAPPIST-1b lies somewhere between these end-members (for example, Fig. \ref{fig:planets}, middle), with an iron core, rock mantle, mantle of high pressure polymorphs, and a surface layer of H$_2$O.  This is consistent with more detailed interior models that take into account compression of rock and metal at depth, which suggest that TRAPPIST-1b must contain more than 6 to 8\% H$_2$O \citep{unterborn17}, likely about $25\%$ H$_2$O by mass \citep{wang17}. It seems likely that the planet does have some liquid water, even if it is just a relatively thin layer at the surface \citep{unterborn17}.  Due to its proximity to the star, the effective surface temperature on TRAPPIST-1b, $T_\mathrm{eff}=400$ K.  Thus, the planet must be in a runaway greenhouse state.

\begin{figure}
\centering 
\includegraphics[width=22pc]{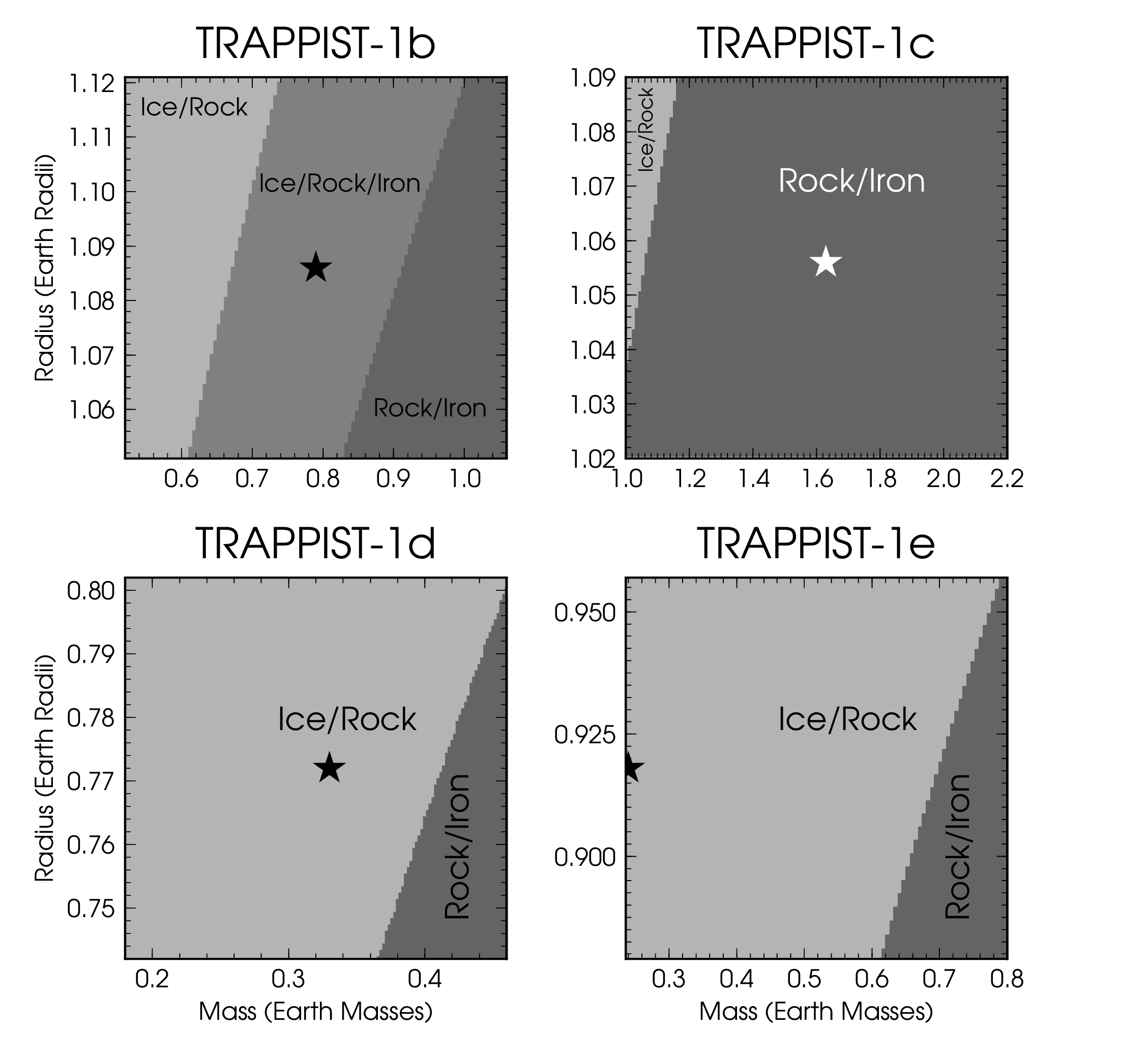}
\caption{Boundaries between ice/rock, ice/rock/iron, and rock/iron compositions as a function of planetary mass and radius for the four innermost TRAPPIST-1 planets. Star indicates the location of the best-fit mass and radius from \citet{wang17}. For TRAPPIST-1b (top, left), gray regions show pairs of mass and radius for which the planet has a four-layer structure, similar to the ``mean density'' case illustrated in the top row of Fig. \ref{fig:planets}.  For TRAPPIST-1c  (top, right), the vast majority of pairs of mass and radius values indicate a rock/iron composition.  For planets d and e (bottom left and right), an ice/rock composition is most likely; structures completely devoid of H$_2$O are indicated only if each planet has a mass close to the highest permitted value and a radius close to the minimum permitted value. \label{fig:four-layer-b} \label{fig:four-layer-c} \label{fig:four-layer-d} \label{fig:four-layer-e}}
\end{figure}

The top, left panel of Fig. \ref{fig:four-layer-b} illustrates how the interior structure and composition of TRAPPIST-1b varies as a function of planetary mass and radius.  The figure depicts the range of masses and radii for which TRAPPIST-1b is composed of ice, rock, and metal in a four-layered interior structure similar to the middle panel in the top row of Fig. \ref{fig:planets}.  At smaller masses and larger radii, TRAPPIST-1b is composed of ice and rock; for larger masses and smaller radii, the planet is made of rock and iron. If the error bars on the mass for TRAPPIST-1b were reduced to $\pm 0.1M_{\oplus}$, it would be possible to determine which of the three compositions best matched the planet's true composition.  

We found that the equilibrium mantle temperature for TRAPPIST-1b was above the rock solidus temperature, suggesting that the rock component of TRAPPIST-1b can be partially molten.  If TRAPPIST-1b has a thick mantle of solid hpp ice, buoyant melt will rise to the top of the mantle, where it can erupt at the rock/ice boundary.  This will drive melting at the base of the high-pressure ice polymorph mantle, driving ice melting and the rise of buoyant water through the hpp ice mantle \citep{barr01}.  

The surface heat flux from tidal heat for TRAPPIST-1b, $F_\mathrm{tidal}=2.68$ W/m$^2$, is twice that of Jupiter's moon Io \citep{spencer00, veeder04}, which is one of the most volcanically active planetary bodies in our Solar System \citep{schubert04}.  Thus, we expect TRAPPIST-1b to be volcanically active, but any activity would be concentrated at the ice/rock boundary, or at the bottom of a thick surface ocean.  This raises the possibility of habitable regions inside TRAPPIST-1b in the vicinity of, for example, hydrothermal vents at the ice/rock boundary, as has been suggested for Jupiter's moon Europa \citep{vance07}. However, the global energy flux on TRAPPIST-1b is about three times larger than the runaway greenhouse limit for the planet, indicating that the surface temperature on the planet may be too high to support life.

\subsection{Planet c}

TRAPPIST-1c has the highest mean density among the planets in the system, with an estimated $\bar{\rho}=7642$ kg/m$^3$.  The simplest permitted structure with such a high density is a large iron core, with a rock mantle, and no H$_2$O layer at the surface (see middle panel in the second row of Fig. \ref{fig:planets}).  This structure is consistent with other interior models, which show that the planet is roughly 50\% iron by mass \citep{wang17}, and less than 6 to 8 \% H$_2$O by mass \citep{unterborn17}.  Figure \ref{fig:planets} also illustrates the two extreme end-member interior structures that were possible given the large error bars on the mean density for TRAPPIST-1c.  Taking the highest possible density for TRAPPIST-1c, the simplest interior structure was essentially a massive iron core with a thin layer of rock at the surface.  The other extreme possibility was a rock-rich planet with no metal core and a mantle of H$_2$O (see right panel of Fig. \ref{fig:planets}).  

The top, right panel of Fig. \ref{fig:four-layer-c} illustrates the pairs of mass and radius for TRAPPIST-1c where a rock/iron composition is permitted, and the small region of parameter space where TRAPPIST-1c could be composed of rock and H$_2$O.  Further improvements to the mass measurement of TRAPPIST-1c could help distinguish between these models; if we knew the mass to within $\pm 0.5 M_{\oplus}$, we could determine whether TRAPPIST-1c might harbor a layer of H$_2$O near its surface.  Consistent with others' interior models, we found that H$_2$O-bearing structures for TRAPPIST-1c are possible only if its density is close to the lower limit permitted by the error bars.  

The equilibrium mantle temperature in the rock layer inside TRAPPIST-1c was $T_{eq}=1666$ K, above the solidus temperature for rock, and a surface tidal heat flux of 1.3 W/m$^2$, comparable to Jupiter's moon Io.  Because it is unlikely that TRAPPIST-1c has a surface layer of H$_2$O, it is possible that TRAPPIST-1c may have surface eruptions of silicate magma.  The high heat flow may be large enough to be observed in infrared wavelengths. The heat flow on Io is known to be spatially and temporally variable as new volcanic vents form, or eruptions occur from previously dormant calderas (e.g., \citealt{veeder12b}).  The surface of Io also harbors numerous lava lakes whose surfaces periodically rupture and founder, resulting in momentary increases in the global heat flow \citep{veeder12b}.  Thus, as instrumentation improves, searching for excess thermal emission from the TRAPPIST-1 system may provide an avenue for the discovery of volcanism in the system.

In the mean density case, the iron core had a radius of $\sim4850$~km which is larger than the iron core of the Earth (radius $\sim3500$~km). This implies that planet c might have a strong magnetic field protecting the surface from stellar wind and flare erosions. However, due to the close-in orbit, the global energy flux is almost twice as large as the runaway greenhouse limit, which probably results in an uninhabitable surface.

\subsection{Planet d}

Planet d is much smaller than its inner two neighbors, with a mass $M=0.33 \pm 0.15M_{\oplus}$, and radius $R=0.772 \pm 0.030R_{\oplus}$.  The third row of Fig. \ref{fig:planets} illustrates the range of possible interior structures permitted by the mass and radius estimates.  In the high-density limit, the planet could be composed solely of rock and iron, with an iron core of radius 2820 km.  In the low-density limit, the planet could have no iron and be composed of rock and H$_2$O (right panel of Fig. \ref{fig:planets}).  If planet d has any significant water content, it would be concentrated near the surface of the planet and would form a liquid water ocean at the surface because $T_\mathrm{eff} > 273$ K.  

Figure \ref{fig:four-layer-d} illustrates the range of possible compositions for TRAPPIST-1d.  For the vast majority of masses and radii, a significant H$_2$O component was required to match the observations.  If the mass of the planet is close to its maximum value, and radius close to the minimum permitted value, structures with no H$_2$O can match the observations. Determining the mass of TRAPPIST-1d to $\pm 0.07 M_{\oplus}$ would shed light on whether TRAPPIST-1d has significant water content.

The equilibrium interior temperature in the rocky component of TRAPPIST-1d $T_{eq}=1618$~K was extremely close to the solidus temperature for rock, indicating that the mantle could be partially molten, but certainly low-viscosity and convective.  Because TRAPPIST-1d likely has a thick layer of H$_2$O on the surface, any melt extraction from the mantle would result in melting of the solid ice mantle.  With an effective surface temperature $T_\mathrm{eff}=288$~K, TRAPPIST-1d is likely to be covered by a global water ocean.  With a global heat flux from tidal heating of 160~mW/m$^2$, TRAPPIST-1d experiences a tidal heat flow twenty times the mean heat flow of the Earth, which provides an abundance of geothermal energy to drive chemistry in its surface ocean. 

We only found an equilibrium temperature for the high-density case for high orbital eccentricities. In the cases of lower eccentricity, the tidal heating rate was not high enough to drive convection in the body, hence no equilibrium temperature can be found. In other words, this indicated that the planet may not experience significant tidal heating, and/or that a more sophisticated tidal model may be needed for these cases to determine the balance between tidal heating and convection. The error bars for the eccentricity of planet d permitted the possibility that the planet's orbit is almost circular, implying insignificant tidal heating.

According to the calculations of \citet{kopparapu17}, planet d is in the runaway greenhouse state. In contrast, we found that the global energy flux is lower than the runaway greenhouse limit, if the planetary albedo is $\gtrsim 0.3$. However, \citet{kopparapu17} uses a climate model for Earth-like bodies, but our model \citet{pierrehumbert10} took into account the mass of the planet (through its surface gravity), which was significantly smaller than the Earth ($0.33 \pm 0.15 M_\oplus$). Because the planet was very close to the runaway greenhouse limit (if not exceeded it), liquid water on its surface was only possible if the planet was tidally locked to the star. Because of the synchronization, a high albedo cloud above the substellar point can reflect most incoming radiation \citep{yang13,kopparapu16}.

\subsection{Planets e, f, and g}

As a group, TRAPPIST-1e, f, and g have somewhat similar masses, and radii, yielding similar mean densities.  Each of the three planets is less massive than Earth but with a radius comparable to that of the Earth, yielding mean densities in between those for ice and rock, similar to the mean densities of  outer planet satellites and dwarf planets in our Solar System, including Pluto \citep{stern15}, Enceladus \citep{porco06}, and Ganymede \citep{pappalardo04}.  

The mass of planet e is quite uncertain, $M=0.24^{+0.56}_{-0.24}M_{\oplus}$: the error bars encompass $M=0$.  As a result, the mean density of planet e is also not well-constrained: $\bar{\rho}=1713^{+4790}_{-413}$ kg/m$^3$ \citep{wang17}.  In the context of our assumed densities for ice, rock, and iron, the minimum density permitted for planet e was $\bar{\rho}_\mathrm{min}=1300$ kg/m$^3$, the density of the high-pressure ice polymorphs.  With this density, the minimum permitted mass for planet e is $M=0.206 M_{\oplus}$, with a radius of $R=0.957R_{\oplus}$.  Assuming a low-density core composed of Prinn-Fegley rock with $\rho_\mathrm{r}=3300$ kg/m$^3$, TRAPPIST-1e could harbor a small rock core, even though its minimum density was comparable to $\rho_\mathrm{hpp}$ (see Table \ref{table:masses}).  The fourth row of Fig. \ref{fig:planets} illustrates the end-member interior structures, ranging from a water-free planet to one composed of almost 100\% H$_2$O (right).  With an effective surface temperature $T_\mathrm{eff}=251$~K \citep{gillon17} below the melting point of water ice at low pressures (273~K), it is possible that TRAPPIST-1e has a solid ice surface. The relative sizes of the liquid water ocean and polymorph layers depended on the temperature in the interior of the planet, but the maximum thickness of the ice I layer can be constrained by the depth to the 209 MPa phase transition among ice I, liquid water, and ice III \citep{hobbs74}.  

The bottom, right panel of Fig. \ref{fig:four-layer-e} illustrates the pairs of mass and radius for which TRAPPIST-1e had an H$_2$O-rich composition.  If the mass of TRAPPIST-1e is close to the maximum permitted value, the planet could contain only rock and iron, but for the vast majority of mass/radius pairs, a thick ice mantle is predicted.  To rule out a water-free structure, one would have to determine the mass of TRAPPIST-1e to within $\sim 0.5 M_{\oplus}$.  

Planets f and g have masses and radii similar to planet e, and thus, similar mean densities and interior structures.  Both planets are less dense than TRAPPIST-1e, and so it seems likely that both of the planets have a significant amount of H$_2$O.  Similar to planet e, the mass of TRAPPIST-1f is not well-constrained, and the error bars permit the planet to have a low-density structure with a mean density equal to that of the high-pressure ice polymorphs.  Because there is a thin layer of ice I or water at the surface, the low-density structure could have a very small Prinn-Fegley rock core.  

For planets e and f, we found that their rock mantles would have equilibrium temperatures near the solidus of rock, similar to planet d.  Also their global heat fluxes from tidal heating were $\sim 160$ to 180 mW/m$^2$ similar to planet d. Both planets f and g are far enough from their parent star to have $T_\mathrm{eff} < 273$ K \citep{gillon17}, suggesting that their surfaces could be solid H$_2$O ice. Given that the tidal heat flow is about twenty times stronger than Earth, it seems likely that both planets could harbor liquid water oceans, perched between the surface layer of ice I and underlain by a mantle of hpp ices, similar to the jovian moons Ganymede and Callisto \citep{schubert04}.  

We were not able to determine an equilibrium temperature for planet g, nor for the low-density case of planet f. This could be because tidal forces from the star are too small to significantly deform the rock cores of these planets, implying that tidal dissipation in the rock mantles of each planet is negligible.  Tidal dissipation could generate heat in the ice mantles of the planet, but a more sophisticated model for $k_2$ would be needed to determine magnitude of the heat and its partitioning between the hpp ice mantle and ice I shell.

\subsection{Planet h}
The outermost planet, TRAPPIST-1h, has a mass of only $M=0.086 \pm 0.084M_{\oplus}$, and a radius of $0.715 R_{\oplus}$ \citep{wang17}.  Given our assumed densities of rock and ice, TRAPPIST-1h, with a mean density $\bar{\rho}=1299$ kg/m$^3$, should contain only H$_2$O, and no rock or metal. 

Although it was permitted in the context of our assumptions, we thought it unlikely that TRAPPIST-1h would be composed of pure water ice.  The vast majority of ice-rich bodies in the outer solar system contain some fraction of rock (one possible exception being Saturn's small moon Tethys, which may not be primordial \citep{SalmonCanup2017}).  In our solar system, almost all of the outer planet satellites and dwarf planets contain some rock, and from a formation standpoint, it seems unlikely that TRAPPIST-1h is completely devoid of rock. For this reason we determined interior structures using a low-density mineral assemblage commonly used as a proxy for the rocky component of outer Solar System satellites: Prinn-Fegley rock, for the planet, and a planetary mean density of 1300~kg/m$^3$.  

We found that a model TRAPPIST-1h with a density $\bar{\rho}_{mean}$ could harbor a core of rock 1030 km in radius, with the remainder of the planet composed of H$_2$O in various phases (see right panel in the last row of Fig. \ref{fig:planets}).  If TRAPPIST-1h had a density close to its maximum permitted value, $\bar{\rho}_{max}=3149$ kg/m$^3$, it could be composed of almost pure rock, with only $\sim100$ km of H$_2$O near the surface (see left panel in the last row of Fig. \ref{fig:planets}).  With its low effective surface temperature, $T_\mathrm{eff}=169$~K \citep{gillon17}, planet h undoubtedly has a solid ice surface.  Similar to our null result for TRAPPIST-1g, we were not able to determine an equilibrium temperature dictated by the balance of tidal heat generation and transport in TRAPPIST-1h. Further knowledge of the composition and interior structure of TRAPPIST-1g and h, as well as a more sophisticated tidal model, will be required to assess the role of tidal heating in its interior evolution.

\section{Implications for habitability} \label{sect:habitability}
TRAPPIST-1b may contain a large amount of H$_2$O, and is likely volcanically active, but its activity is probably concentrated at the ice/rock boundary deep inside the planet. Due to its proximity to the star, the planet is very likely to be in a runaway greenhouse state \citep{kopparapu17, bolmont17, bourrier17}.  TRAPPIST-1c is not likely to have much water or ice, and could have surface eruptions of silicate magma in the style of Io. Detection of excess thermal emission from the TRAPPIST-1 system could provide indirect evidence for surface volcanism on these bodies. 

Using 1-dimensional and 3-dimensional climate models, \citet{gillon17} deduce that the surface temperatures of planets e, f, and g are suitable for harboring water oceans on their surface. However, using the runaway greenhouse limit of \citet{pierrehumbert10}, we found that planet d, too, may be habitable, if its albedo is $\gtrsim 0.3$. Planet d might be covered by a global water ocean that can provide a favorable environment for the appearance of life. This conclusion supports the finding of \citet{vinson17}, who calculate the surface temperature of the TRAPPIST-1 planets due to stellar irradiation and tidal heating due to the circularization of the orbits, and found that planets d, e, and f might be habitable.

Planets e, f, g, and h are likely to have icy surfaces with liquid oceans underneath. According to \citet{checlair17}, however, if the planets are tidally locked and have an active carbon cycle, then they should not have solid ice surfaces. The three-dimensional climate models of \citet{turbet17} show that TRAPPIST-1e is able to sustain its water content in liquid phase regardless of its atmospheric composition, provided that the planet is synchronously locked and abundant in water. In the atmospheres of planets f and g, a few bars of carbon dioxide is needed to facilitate the melting of ice on their surfaces. Most studies find that planet e is most likely to be habitable among the TRAPPIST-1 planets. For example, according to \citet{wolf17}, even a thin atmosphere could be enough to sustain temperatures of $\sim 280$~K locally around the substellar point, melting $\sim 13$~\% of the surface ice. \citet{vinson17} also find that including tidal heating to the energy budget, the surface temperature of planet e is around 280~K.

According to the tidal heating calculations of \citet{papaloizou17}, planet e, and in most of the investigated cases, also planet f, could be both in the conservative habitable zone and in the tidal habitable zone. The tidal habitable zone indicates an intermediate level of tidal heating flux on the planetary surface (between 0.04 and 2~W/m$^2$) which could drive plate tectonics required for the carbonate-silicate cycle that stabilizes the atmospheric temperature of the planet \citep{barnes09}. \citet{papaloizou17} also find that planet d orbits in the optimistic habitable zone, and its tidal heating is sufficient to be in the tidal habitable zone, too.

\citet{bourrier17} find that during the assumed 8 Gyr lifetime of the system, planets b through f may have lost, together, more than 20 Earth oceans worth of water due to hydrodynamic escape. However, the estimation is highly uncertain because of the wide possible ranges in planetary masses, and their values should be considered upper limits for hydrodynamical water loss. Magnetic interaction with the host star, however, can result in additional loss of water \citep{dong17}.

In any case, on planets b, e, f, g, and h, life might appear in the tidally heated (subsurface) ocean close to hydrothermal vents. Although our tidal heating model did not constrain equilibrium temperatures for planets g and h, they may still experience tidal heating. Our model might not be suitable for low heating rates: tidal heating can also be balanced by conduction alone if the planet's mantle is too viscous to convect.  We have also neglected planet--planet tidal interactions which may significantly contribute to the tidal heating in each planet, because they orbit in a close mean motion resonant chain. In addition, using tidal evolution models, \citet{barnes17} find that planets in the habitable zone of late type M dwarfs are very likely to enter a 1:1 spin--orbit resonance in less than 1 Gyr. If the planet's rotation is fast, then the orbital eccentricity is expected to grow slowly, which will lead to somewhat stronger tidal heating. 

\citet{lingam2017b} investigate the probability of habitability of three TRAPPIST-1 planets, e, f, and g, orbiting inside the circumstellar habitable zone. By calculating the likelihood functions for habitability based on surface temperature and atmospheric loss of the planets (due to hydrodynamic escape and stellar winds), they find that for planet e, which gave the most promising results, the likelihood function is at least one to two orders of magnitude lower than in the case of Earth.

Any form of life is threatened however, by strong stellar activity. Analysis of K2 observations of TRAPPIST-1 showed frequent flares which present a serious threat for planetary atmospheres \citep{vida17,garraffo17}. Strong magnetic fields are required to protect the atmosphere from erosion, and life from high energy radiation.

\section{Summary}
The seven planets of the TRAPPIST-1 system have been the focus of intense observation and debate since the discovery of the system.  Several of the planets are roughly Earth-sized, and their mean densities are not dissimilar from the terrestrial planets and large icy satellites in our Solar System.  Moreover, the effective surface temperatures of each of the planets are relatively modest, suggesting solid or water-covered surfaces.  

Each of the seven TRAPPIST-1 planets are quite close to the parent star, as indicated by their orbital periods which range from 1.5 to about 20 days.  The planets also occupy a mean-motion resonance \citep{gillon17} and have non-zero orbital eccentricities.  This implies that the planets experience tidal heating, which can be a significant energy source.  

Here, we have constructed interior models for the planets using constant densities for ice, rock, and metal. The masses and radii of each of the planets are not well-constrained and have large uncertainties; for this reason, we applied our model using the mean density of each planet and also the extrema obtained from the error bars of observations. Within the range of estimated masses and radii for planets b, c, d, and e, each planet could be composed of solely iron and rock (our ``high-density'' cases).  If the true densities of planets b, d, and e are close to the mean values implied by observations, each of these planets could have a layer of H$_2$O near the surface.  The large density of planet c points toward a rock/iron composition, H$_2$O is permitted only if the true density is closer to the lower limit of the estimated value.  

Planets b and c likely harbor magma oceans in their interiors. The surface heat fluxes from tidal heat range from hundreds of milliwatts per meter squared (about twenty times higher than the Earth's heat flow) to watts per meter squared, comparable to the heat flow estimated for Jupiter's volcanically active moon, Io. Further work is required to reconcile the seemingly water-rich composition of planet b with its proximity to the TRAPPIST-1 star.  The majority of permitted compositions for planet c indicate that the body is water-poor, raising the possibility that melt extracted from the magma ocean could erupt onto the surface.  Further observations of the TRAPPIST-1 system may be able to detect the thermal or chemical signature of this activity, if present.

Of the seven planets in the TRAPPIST-1 system, we consider planets d and e to be the most likely candidates for habitable environments.  Planet d is likely to be covered by a global water ocean, and can be habitable if its albedo is $\gtrsim$~0.3. Planetary albedos may be estimated from photometric measurements during occultations \citep[see e.g.][]{rowe08} which may give an additional constraint on the habitability of TRAPPIST-1d. As discussed in Section \ref{sect:habitability}, planet e is likely to have liquid water on its surface, too, according to climate models \citep{turbet17,wolf17}.

Planets f, g, and h are undoubtedly H$_2$O-rich, with thick mantles of ice, possible subsurface liquid water oceans, and solid surfaces composed of ice I, the low-density phase of ice present in terrestrial glaciers and ice sheets.  It is worth noting that the present estimates of the mean density of planet h permit a 100\% H$_2$O composition, but this seems unlikely given that even the most ice-rich bodies in our Solar System (for example, the Kuiper Belt Objects) have a significant amount of rock in their interiors \citep{barr16}.  The geodynamical evolution of the outer planets of TRAPPIST-1 should be studied in more detail with a layered tidal and thermal model that can self-consistently determine the amount of tidal heating and heat transport by convection in both the rock and ice mantles.

Given the uncertainties in the compositions of each of the planets, we have employed a relatively simple geophysical model to determine the balance between tidal heat production and transport in each planet.  Our approach provides more realistic values for $Im(k_2)$ than a constant-$Q$ approach because we take into account the feedback between the temperature of the planet. We used a tidal model that assumes each planet is a homogeneous sphere, but mimic the effect of the presence of multiple materials by determining an effective viscosity and rigidity for the planet based on the viscosities and rigidities of its component materials. We focused our efforts on determining the temperature of the rock mantle in each planet by balancing heat produced with the amount of heat transported by solid-state convection in the rock layer of each planet. When updated masses and radii are available, we will be pursuing more detailed calculations of tidal heating using multi-layered models and numerical techniques.

Further observations of the TRAPPIST-1 system are required to provide better constraints on the composition and interior structure of each planet. When improved compositional constraints are obtained, more sophisticated modeling techniques that calculate tidal dissipation and energy transport in a multi-layered planet can provide improved estimates of the tidal heat flux and mantle temperature.  Our knowledge of the compositions of the four interior planets, chiefly, their water content, can significantly be improved if their masses can be determined to within $\sim 0.1$ to $0.5 M_{\oplus}$, which will require further observations of the TRAPPIST-1 system.  

\section*{Acknowledgements}

We thank the anonymous referee for a thorough review of the manuscript and helpful comments. We also thank Dr. L\'aszl\'o Moln\'ar for initiating this collaboration, and Dr. Jessica Irving, Dr. Tess Caswell, and Dr. \'Akos Kereszturi for helpful conversation. VD has been supported by the Hungarian National Research, Development and Innovation Office (NKFIH) grants K119993, K-115709 and GINOP-2.3.2-15-2016-00003.


\end{document}